\documentclass[prb,preprintnumbers,amsmath,amssymb,floatfix]{revtex4}

\usepackage{graphicx}
\usepackage{subfigure}
\usepackage{dcolumn}

\begin{document}
\title{ Logical operation of one-dimensional photonic crystal based on series and parallel connection}
\author{Xiang-Yao Wu$^{a}$ \footnote{E-mail: wuxy2066@163.com},
 Ji Ma$^{a}$, Xiao-Jing Liu$^{a}$\\ Xiang-Dong Meng$^{a}$, Yu Liang$^{a}$, Hong Li$^{a}$ and Si-Qi Zhang$^{a}$}
 \affiliation{a. Institute of Physics, Jilin Normal
University, Siping 136000 China}
\begin{abstract}
In this paper, we have proposed the compound structure of
one-dimensional photonic crystal (PC), which includes series
connection and parallel connection PC. We have studied the
transmission characteristics of series connection and parallel
connection PC, and obtained some new results. In addition, we have
proved the series connection one-dimensional PC can realize the
logical AND operation, and the parallel connection one-dimensional
PC can realize the logical OR operation. The compound structure of
one-dimensional PC can design more new type structure optical
devices, and will provide the basic for designing quantum
computer.

\end{abstract}
\maketitle

 \vskip 8pt
 {\bf 1. Introduction} \vskip 8pt
E. Yablonovitch and S. John had pointed out that the behavior of
photons in 1987. It can be changed when propagating in the
material with periodical dielectric constant, and termed such
material PC [1,2]. PC important characteristics are: photon band
gap, defect states, Light localization and so on. These
characteristics make it able to control photons, so it may be used
to manufacture some high performance devices which have completely
new principles or can not be manufactured before, such as
high-efficiency semiconductor lasers, right emitting diodes, wave
guides, optical filters, high-Q resonators, antennas,
frequency-selective surface, optical wave guides and sharp bends
[3-4], WDM-devices [5-6], splitters and combiners [7]. optical
limiters and amplifiers [8-10]. The research on PC will promote
its application and development on integrated photoelectron
devices and optical communication. To investigate the structure
and characteristics of band gap, there are many methods to analyze
PC including the plane-wave expansion method [11], Green¡¯s
function method, finite-difference time-domain method [12-14] and
transfer matrix method [15-20].

In this paper, we have proposed the compound structure
one-dimensional PC, which include series connection, parallel
connection compound structure PC. We have studied their
transmission characteristics and given the relation of logical
operation. The work frequency of light signal is taken as the
defect mode frequency of PC. When the PC has the defect mode,
transmissivity $T=1$ corresponds to logical $1$. When the PC has
no defect mode, transmissivity $T=0$ corresponds to logical $0$.
We can find the series connection one-dimensional PC can realize
the AND operation, and the parallel connection one-dimensional PC
can realize the OR operation. The compound structure PC will help
to design more new type structure optical devices, and provide the
basic for designing quantum computer.

 \vskip 8pt
 {\bf 2. Transfer matrix and transmissivity of one-dimensional
PC} \vskip 8pt

For one-dimensional PC, the calculations are performed using the
transfer matrix method [21], which is the most effective technique
to analyze the transmission properties of PC. For the medium layer
$i$, the transfer matrices $M_i$ for $TE$ wave is given by [21]:
\begin{eqnarray}
M_{i}=\left(%
\begin{array}{cc}
 \cos\delta_{i} & -i\sin\delta_{i}/\eta_{i} \\
 -i\eta_{i}sin\delta_{i}
 & \cos\delta_{i}\\
\end{array}%
\right),
\end{eqnarray}
where $\delta_{i}=\frac{\omega}{c} n_{i} d_i cos\theta_i$, $c$ is
speed of light in vacuum, $\theta_i$ is the ray angle inside the
layer $i$ with refractive index $n_i=\sqrt{\varepsilon_i \mu_i}$,
$\eta_i=\sqrt{\varepsilon_i/\mu_i} cos\theta_i$,
$cos\theta_i=\sqrt{1-(n^2_0sin^2\theta_0/n^2_i)}$, in which $n_0$
is the refractive index of the environment wherein the incidence
wave tends to enter the structure, and $\theta_0$ is the incident
angle.

The total transfer matrix $M$ for an $N$ period structure is given
by:
\begin{eqnarray}
\left(%
\begin{array}{c}
  E_{1} \\
  H_{1} \\
\end{array}%
\right)&=&M_{B}M_{A}M_{B}M_{A}\cdot\cdot\cdot M_{B}M_{A}\left(%
\begin{array}{c}
  E_{N+1} \\
  H_{N+1} \\
\end{array}%
\right)
\nonumber\\&=&M\left(%
\begin{array}{c}
  E_{N+1} \\
  H_{N+1} \\
\end{array}%
\right)=\left(%
\begin{array}{c c}
  A &  B \\
 C &  D \\
\end{array}%
\right)
 \left(%
\begin{array}{c}
  E_{N+1} \\
  H_{N+1} \\
\end{array}%
\right),
\end{eqnarray}
where
\begin{eqnarray}
M=\left(%
\begin{array}{c c}
  A &  B \\
 C &  D \\
\end{array}%
\right),
\end{eqnarray}
with the total transfer matrix $M$, we can obtain the
transmissivity $T$, it is
\begin{eqnarray}
T=|\frac{E_{N+1}}{E_{1}}|^2=|\frac{2\eta_{0}}{A\eta_{0}+B\eta_{0}\eta_{N+1}+C+D\eta_{N+1}}|^2.
\end{eqnarray}
Where $\eta_{0}=\eta_{N+1}=\sqrt{\frac{\varepsilon_0}{\mu_0}}
\cos\theta_0$. By the Eqs. (1) and (4), we can calculate the
transmissivity of one-dimensional PC. \vskip 8pt
 {\bf 3. Series connection and parallel connection one-dimensional PC transmissivity} \vskip 8pt

Two or multiple one-dimensional PC can be connected by optical
fiber, which can be designed series connection, parallel
connection compound structure one-dimensional PC. The FIGs. 1 and
2 are series connection and parallel connection one-dimensional PC
structures, respectively. The $PC_1$ and $PC_2$ are the
one-dimensional PC, $E_{in}$ is the input electric field
intensity, and $E_{out1}$ and $E_{out2}$ are the output electric
field intensity of $PC_1$ and $PC_2$.

\begin{figure}[tbp]
\includegraphics[width=8 cm]{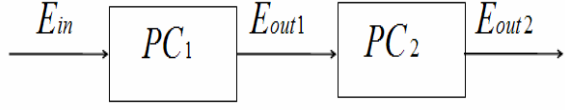}
\caption{The series connection structure one-dimensional PC.}
\end{figure}
\begin{figure}[tbp]
\includegraphics[width=8 cm]{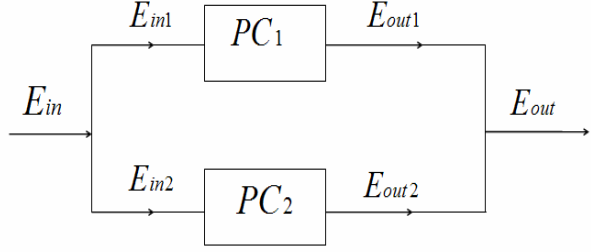}
\caption{The parallel connection structure one-dimensional PC.}
\end{figure}

(1) The total transmission coefficient $t$ and transmissivity $T$
of series connection one-dimensional PC are
\begin{eqnarray}
t=\frac{E_{out2}}{E_{in}}=\frac{E_{out2}}{E_{out1}}\cdot\frac{E_{out1}}{E_{in}}=t_2\cdot
t_1,
\end{eqnarray}
\begin{eqnarray}
T=|t|^2,
\end{eqnarray}
where $t_2={E_{out2}}/{E_{out1}}$ and $t_1={E_{out1}}/{E_{in}}$
for the transmission coefficients of $PC_1$ and $PC_2$.

Similarly, we can obtain the transmission coefficient for $n$
series connection PC, such as $PC_1$, $PC_2$, $\cdot\cdot\cdot$,
$PC_n$ series connection transmission coefficient is
\begin{eqnarray}
t=t_n\cdot t_{n-1}\cdot\cdot\cdot t_1.
\end{eqnarray}

(2) The total transmission coefficient $t$ and transmissivity $T$
of parallel connection one-dimensional PC are
\begin{eqnarray}
t_{\pm}=\frac{E_{out}}{E_{in}}=\frac{E_{out1}\pm
E_{out2}}{E_{in}}=\frac{t_1E_{in1}\pm t_2E_{in2}}{E_{in}},
\end{eqnarray}
when $E_{out1}$ and $E_{out2}$ phase are the same (opposite), the
numerator of Eq. (8) takes $+$ ($-$).

when $t_1=t_2=t$, there are
\begin{eqnarray}
t_{+}=t_1=t_2=t, \hspace{0.5in} t_{-}=\frac{t(E_{in1}-
E_{in2})}{E_{in}}=t(c_1-c_2),
\end{eqnarray}
where $c_1={E_{in1}}/{E_{in}}$ and $c_2={E_{in2}}/{E_{in}}$.

when $t_1\neq t_2$, there is
\begin{eqnarray}
t_{+}=c_1t_1+ c_2t_2, \hspace{0.5in} t_{-}=c_1t_1- c_2t_2
\hspace{0.1in} (c_1+c_2=1).
\end{eqnarray}
Similarly, the total transmission coefficient of $n$ PC $PC_1$,
$PC_2$, $\cdot\cdot\cdot$, $PC_n$ parallel connection is
\begin{eqnarray}
t_{\pm}&=&\frac{E_{out}}{E_{in}}=\frac{E_{out1}\pm E_{out2}\pm
\cdot\cdot\cdot \pm
E_{outn}}{E_{in}}\nonumber\\
&=&\frac{t_1E_{in1}\pm t_2E_{in2}\pm
\cdot\cdot\cdot \pm t_nE_{inn}}{E_{in}}\nonumber\\
&=&c_1t_1\pm c_2t_2\pm \cdot\cdot\cdot \pm c_nt_n.
\end{eqnarray}

 \vskip 8pt
 {\bf 4. Numerical result}
 \vskip 8pt

In this section, we report our numerical results of compound
structure one-dimensional PC, the PC $PC_1$ and $PC_2$ are
constituted by media $A$ and $B$. The main parameters are: the
medium $A$ refractive indices $n_a=2.45$, thickness $a=469nm$, the
medium $B$ refractive indices $n_b=1.35$, thickness $b=890nm$, the
center frequency $\omega_0=\frac{\pi
c}{(n_aa+n_bb)}=4.01\times10^{14}Hz$, the incident angle
$\theta_0=0$.

Firstly, we study the transmission characteristics of series
connection and parallel connection structure PC, which are
constituted by $PC_1$ and $PC_2$, their structure are $(AB)^{12}$.
In $PC_1$, the medium $B$ thickness $b=890nm$. In $PC_2$, the
medium $B$ thickness $b=560nm$, which are shown in FIGs. 1 and 2.
The series connection and parallel connection structure are
referred to as $PC_1\cdot PC_2$ and $PC_1+PC_2$. From Eqs. (5) and
(6), we can calculate the series connection structure
transmissivity. By Eq. (10), we can calculate the parallel
connection structure transmissivity $|t_+|^2$.
\begin{figure}[tbp]
\includegraphics[width=9 cm]{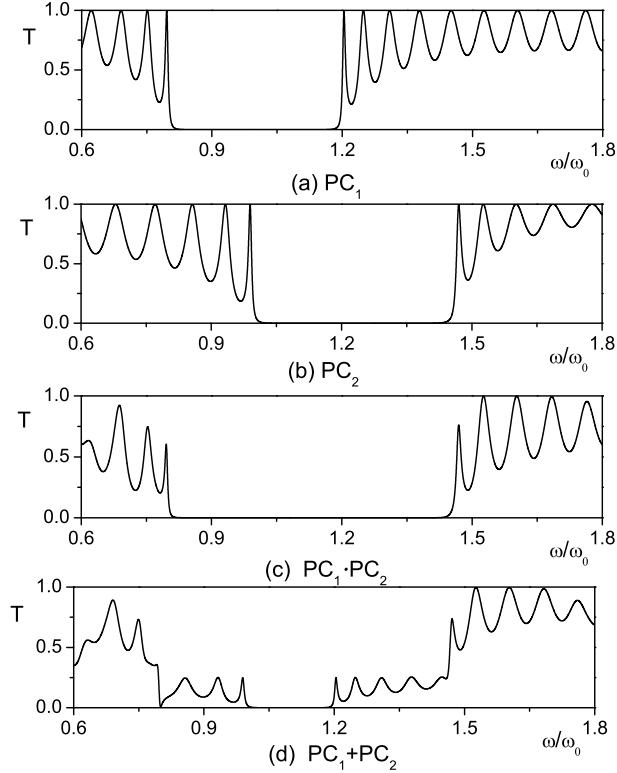}
\caption{The series connection structure transmissivity. (a)
$PC_1$ transmissivity, (b) $PC_2$ transmissivity, (c) series
connection transmissivity, (d) parallel connection
transmissivity.}
\end{figure}

The FIG. 3 (a), (b), (c) and (d) are the transmissivity
corresponding to the structure $PC_1$, $PC_2$, $PC_1\cdot PC_2$
and $PC_1+ PC_2$. From FIG. 3 (a), (b) and (c), we can obtain some
results about the series connection PC: (1) The forbidden band
width of series connection PC becomes more wider, and it is the
union of corresponding forbidden band of part PC $PC_1$ and
$PC_2$, which is similar as the series connection ohm law in
circuit. (2) We can obtain the more wider forbidden band by PC
series connection. (3) With the number of series connection PC
increasing, the total width of forbidden band increase. From FIG.
3 (a), (b) and (d) ($c_1=0.5$). We can find when coefficient
$c_1=0.5$ the forbidden band of parallel connection (FIG. 3 (d))
is the intersection of corresponding forbidden band of $PC_1$ and
$PC_2$.

Secondly, we begin to study the AND logical operation. In FIGs. 4,
5 and 6, we study the AND operation by the series connection
one-dimensional PC $PC_1$ and $PC_2$. The work frequency of light
signal is taken as the defect mode frequency of PC. When the PC
has the defect mode, transmissivity $T=1$ corresponds to logical
$1$. When the PC has no defect mode, transmissivity $T=0$
corresponds to logical $0$. The logical AND operation are: $0\cdot
0=0$, $1\cdot 0=0$ and $1\cdot 1=1$. In FIG. 4, the structures of
$PC_1$ and $PC_2$ both are $(AB)^{12}$ with no defect mode. FIGs.
4 (a), (b) and (c) show the transmissivity corresponding to the
structure $PC_1$, $PC_2$ and $PC_1\cdot PC_2$. $PC_1$ and $PC_2$
both correspond to logic $0$. The series connection $PC_1\cdot
PC_2$ has transmissivity $0$ corresponding to logic AND operation
$0\cdot 0=0$. In FIG. 5, the $PC_1$ is symmetrical structure
$(AB)^{6}(BA)^{6}$ with the defect mode and $PC_2$ structure is
$(AB)^{12}$ without the defect mode. FIGs. 5 (a), (b) and (c) show
the transmissivity corresponding to the structure $PC_1$, $PC_2$
and series connection $PC_1\cdot PC_2$. $PC_1$ corresponds to
logical $1$. $PC_2$ corresponds to logical $0$. The series
connection $PC_1\cdot PC_2$ has transmissivity $0$ corresponding
to logic AND operation $1\cdot 0=0$. In FIG. 6, $PC_1$ and $PC_2$
both are symmetrical structure $(AB)^{6}(BA)^{6}$. They are both
have defect mode. FIGs. 6 (a), (b) and (c) show the transmissivity
corresponding to the structure $PC_1$, $PC_2$ and $PC_1\cdot
PC_2$. At the working frequency, $PC_1$ and $PC_2$ both have
transmissivity $1$ corresponding to the logic $1$. The series
connection $PC_1\cdot PC_2$ has transmissivity $1$ corresponding
to logic AND operation $1\cdot 1=1$.

Finally, we will study the OR logical operation based on the
parallel connection one-dimensional PC $PC_1$ and $PC_2$. The
logical OR operation are: $0+0=0$, $1+0=1$ and $1+1=1$, which
correspond to FIGs. 7, 8 and 9, respectively. In FIG. 7, the
structures of $PC_1$ and $PC_2$ both are $(AB)^{12}$. FIGs. 7 (a),
(b) and (c) show the transmissivity corresponding to the structure
$PC_1$, $PC_2$ and $PC_1+PC_2$ (the parameter $c_1=0.9$), they are
all with no defect mode, correspond to logic $0$, which realize
the logical OR operation $0+0=0$. For FIG. 8 (a), the $PC_1$ is
symmetrical structure $(AB)^6(BA)^6$ with the defect mode
corresponds to logical $1$, For FIG. 8 (b), the $PC_2$ structure
is $(AB)^12$ without the defect mode corresponds to logical $0$.
The FIG. 8 (c) show the transmissivity corresponding to the
parallel connection structure $PC_1+PC_2$ with the defect mode
corresponds to logical $1$, which realize the logical OR operation
$1+0=1$. In FIG. 9 (a) and (b), the $PC_1$ and $PC_2$ are
symmetrical structure $(AB)^6(BA)^6$ with the defect mode
corresponds to logical $1$. FIG. 9 (c) is the transmissivity
corresponding to the parallel connection structure $PC_1+PC_2$
with the defect mode corresponds to logical $1$, which realize the
logical OR operation $1+1=1$.

\vskip 8pt {\bf 5. Conclusion}
 \vskip 8pt

In summary, we have proposed the compound structure
one-dimensional PC, which include series connection and parallel
connection compound structure. We have studied series connection
and parallel connection transmission characteristics and obtained
some new results. In addition, we have proved the series
connection one-dimensional PC can realize the logical AND
operation, and the parallel connection one-dimensional PC can
realize the logical OR operation. We think the other logical
operations can be achieved by other compound structure PC. The
compound structure PC will help to design more new type structure
optical devices, and provide the basic for designing quantum
computer.

\vskip 8pt {\bf 6.  Acknowledgment} \vskip 8pt

This work was supported by the National Natural Science Foundation
of China (no. 61275047), the Research Project of Chinese Ministry
of Education (no. 213009A) and the Scientific and Technological
Development Foundation of Jilin Province (no.20130101031JC).

\begin{figure}[tbp]
\includegraphics[width=9 cm]{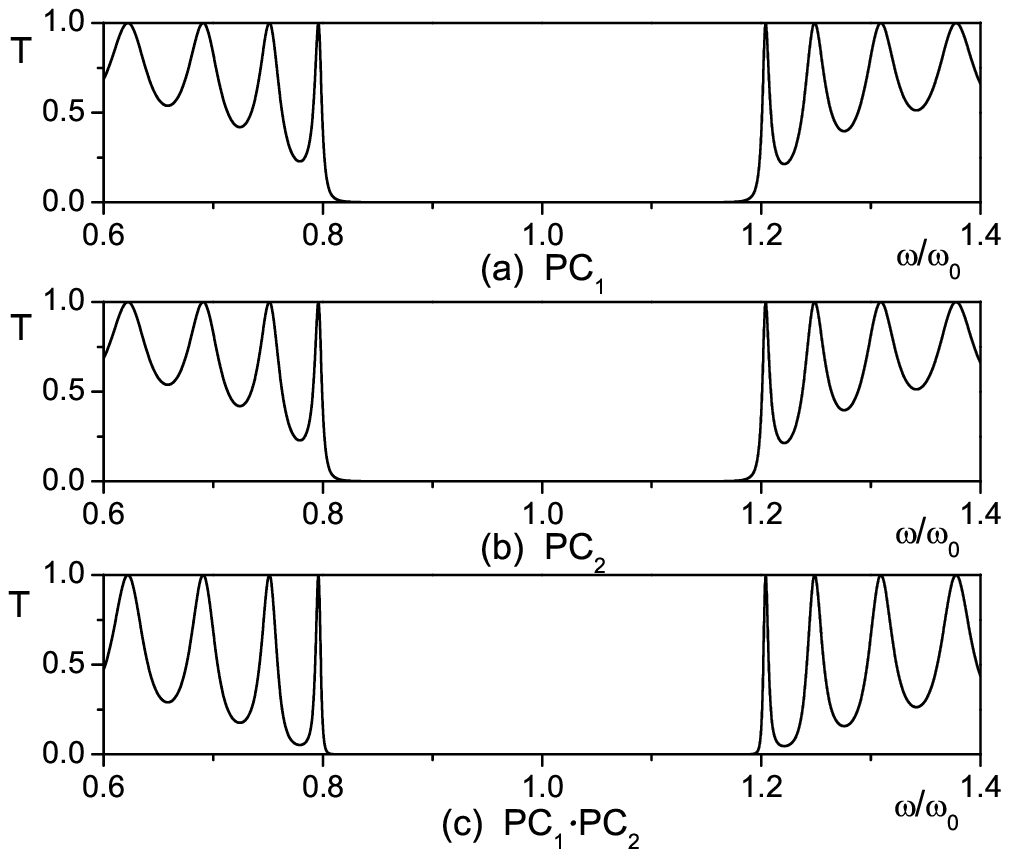}
\caption{The series connection structure transmissivity. (a)
$PC_1$ transmissivity without defect model (logical $0$), (b)
$PC_2$ transmissivity without defect model (logical $0$), (c)
series connection transmissivity without defect model (logical
$0$), realizing logical AND operation $0\cdot0=0$.}
\end{figure}
\begin{figure}[tbp]
\includegraphics[width=9 cm]{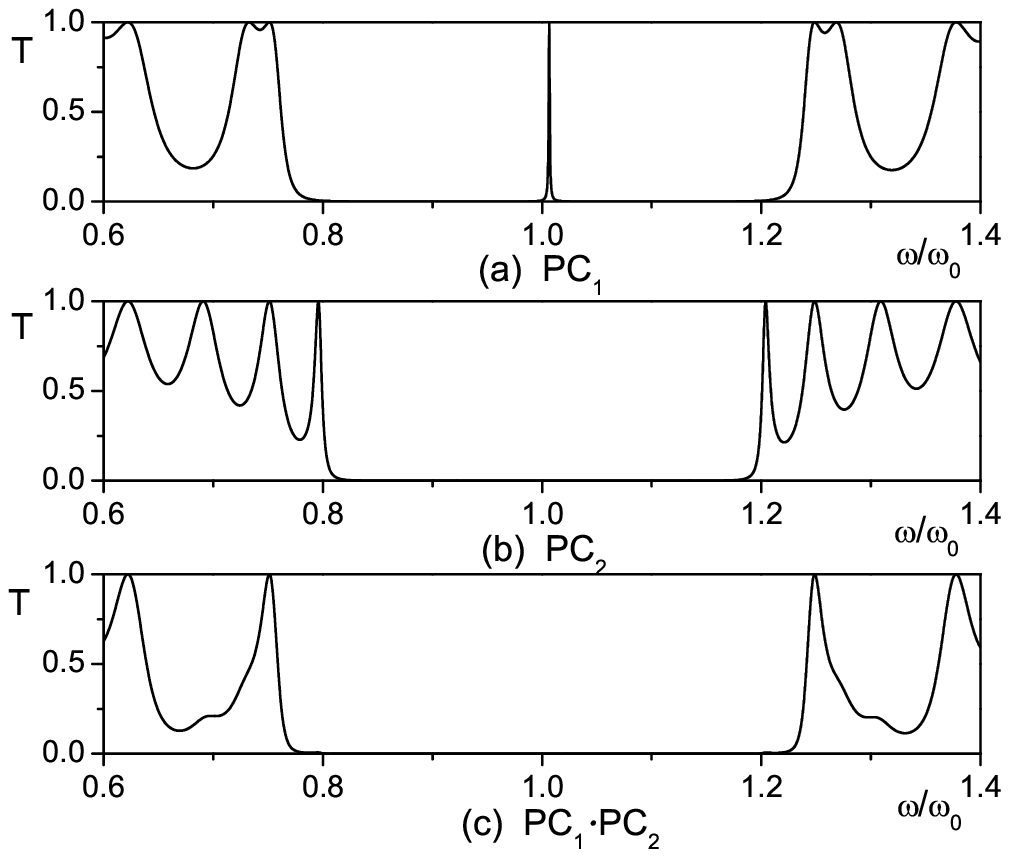}
\caption{The series connection structure transmissivity. (a)
$PC_1$ transmissivity with defect model (logical $1$), (b) $PC_2$
transmissivity without defect model (logical $0$), (c) series
connection transmissivity without defect model (logical $0$),
realizing logical AND operation $1\cdot0=0$.}
\end{figure}
\begin{figure}[tbp]
\includegraphics[width=9 cm]{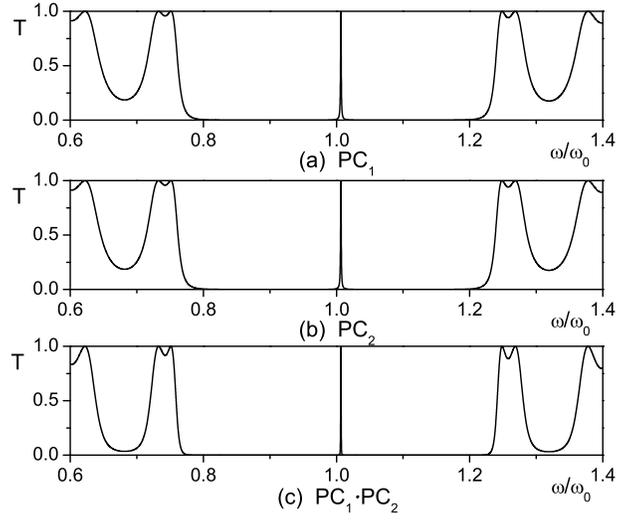}
\caption{The series connection structure transmissivity. (a)
$PC_1$ transmissivity (logical $1$), (b) $PC_2$ transmissivity
(logical $1$), (c) series connection transmissivity (logical $1$,
realizing logical AND operation $1\cdot1=1$.}
\end{figure}
\begin{figure}[tbp]
\includegraphics[width=9 cm]{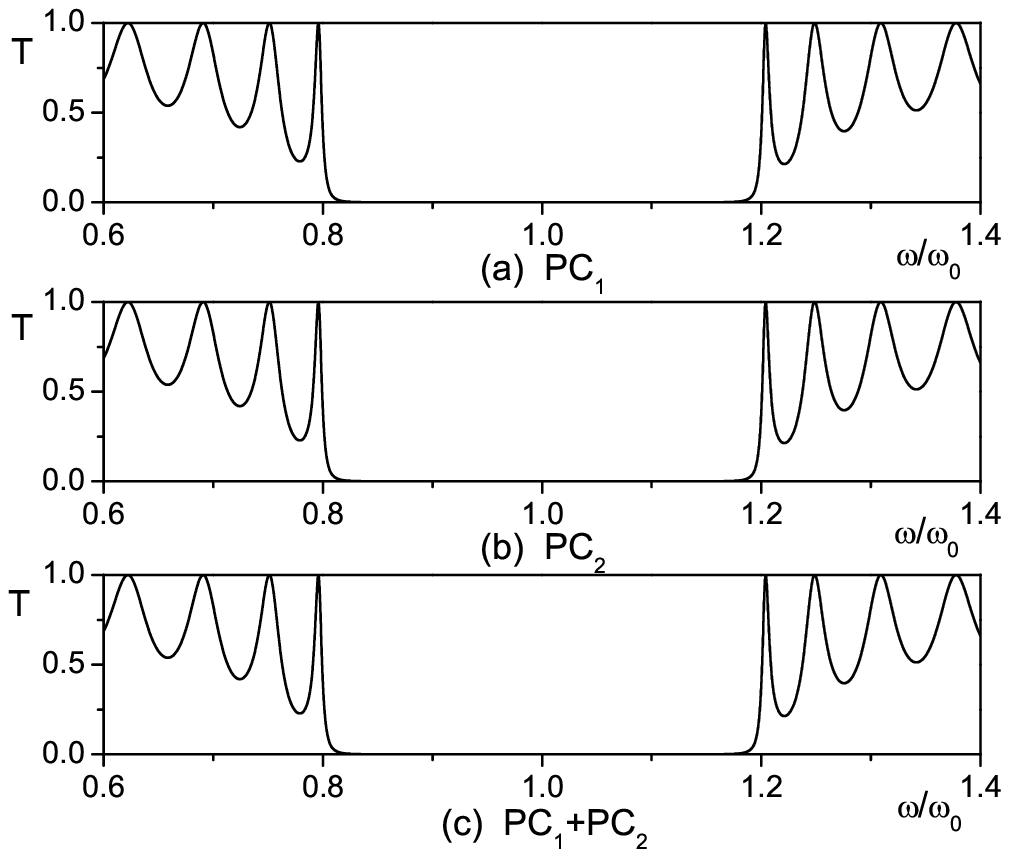}
\caption{The parallel connection structure transmissivity. (a)
$PC_1$ transmissivity without defect model (logical $0$), (b)
$PC_2$ transmissivity without defect model (logical $0$), (c)
series connection transmissivity without defect model (logical
$0$), realizing logical OR operation $0+0=0$.}
\end{figure}
\begin{figure}[tbp]
\includegraphics[width=9 cm]{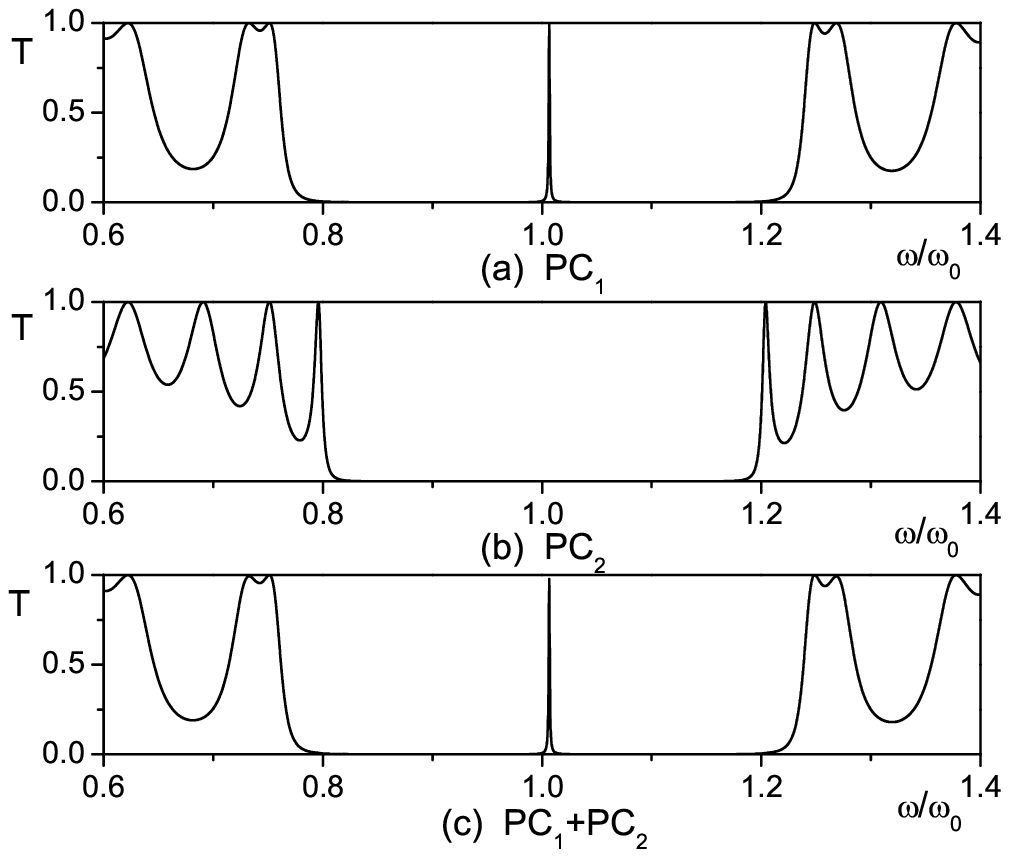}
\caption{The parallel connection structure transmissivity. (a)
$PC_1$ transmissivity with defect model (logical $1$), (b) $PC_2$
transmissivity without defect model (logical $0$), (c) series
connection transmissivity with defect model (logical $1$),
realizing logical OR operation $1+0=1$.}
\end{figure}
\begin{figure}[tbp]
\includegraphics[width=9 cm]{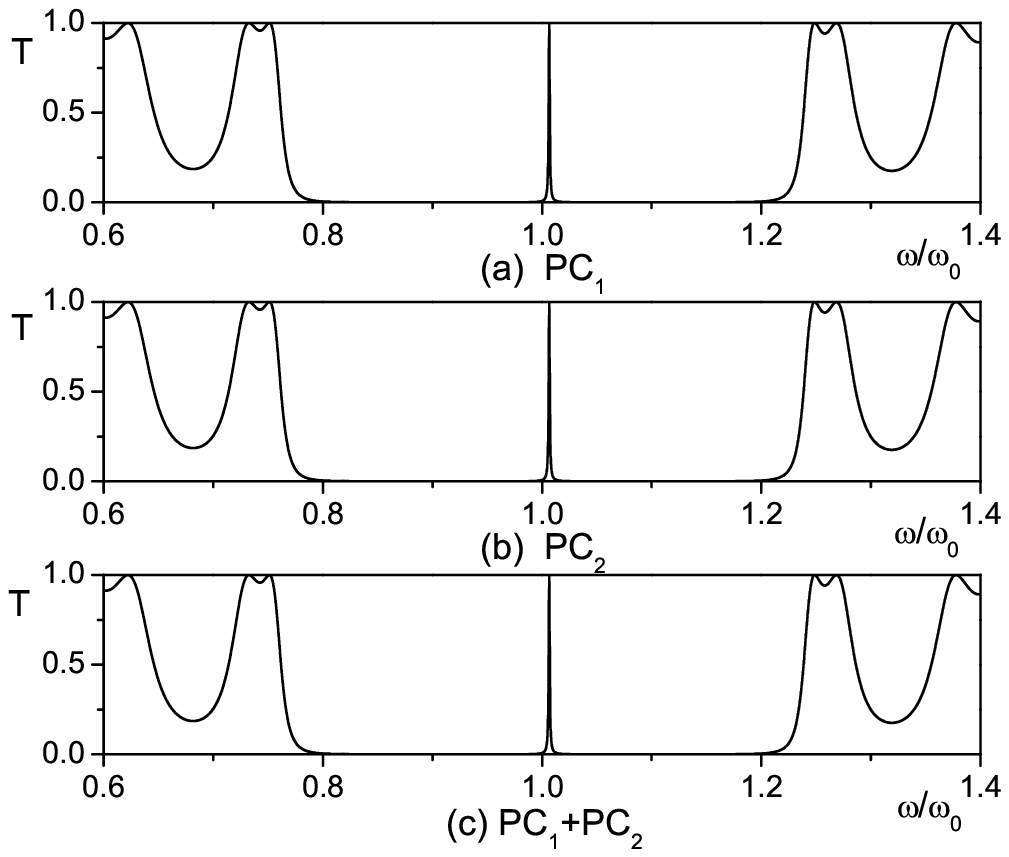}
\caption{The parallel connection structure transmissivity. (a)
$PC_1$ transmissivity with defect model (logical $1$), (b) $PC_2$
transmissivity with defect model (logical $1$), (c) series
connection transmissivity with defect model (logical $1$),
realizing logical OR operation $1+1=1$.}
\end{figure}


\begin{thebibliography}{10}
\bibitem{s1}
E. Yablonovitch, Phys. Rev. Lett. {\bf 58}, 2059 (1987).

\bibitem{s2}
S. John, Phys. Rev. Lett. {\bf 58} 2486 (1987).

\bibitem{s3}
A. Lavrinenko, P.I. Borel, L.H. Frandsen, M. Thorhauge, A. Harpth,
M. Kristensen, T. Niemi, Opt. Express {\bf 12} 234 (2004).

\bibitem{s4}
J. Pu, Y. Yomogida, K. K. Liu, L. J. Li, Y. Iwasa, and T.
Takenobu, Nano Lett., {\bf 12} 4013, (2012).

\bibitem{s5}
S. Fan, P.R. Villeneuve, J.D. Joannopoulos, H.A. Haus, Phys. Rev.
Lett. {\bf 80} 960 (1998).

\bibitem{s6}
A. Ferreira, N. M. R. Peres, R. M. Ribeiro and T. Stauber, Phys.
Rev. B, {\bf 85} 115438, (2012).

\bibitem{s7}
S. Kim, I. Park, H. Lim, Proc. Design of PC splitters/combiners
SPIE {\bf 5597} 129 (2004).

\bibitem{s8}
N. M. R. Peres and Yu. V. Bludov, EPL, {\bf 101} 58002, (2013).

\bibitem{s9}
R. Martinez-Sala, J. Sancho, J. V. Sanchez, V. Gomez, J. Llinares
and F. Meseguer, nature {\bf 378}, 241 (1995).

\bibitem{s10}
T. Cai, R. Bose, G. S. Solomon, and E. Waks, Appl. Phys. Lett.
{\bf 102}, 141118 (2013).

\bibitem{s11}
J. J. Joannopoulos, R. D. Meade, J. N. Winn, PC: molding the flow
of light (Princeton University Press, New Jersey, 1995).

\bibitem{s12}
M. Minkov and V. Savona, Phys. Rev. B {\bf 88}, 081303R (2013).

\bibitem{s13}
K. K. Yee, IEEE Trans. Antennas Propag. {\bf 14}, 302 (1966).

\bibitem{s14}
K. S. Choi, H. Deng, J. Laurat, and H. J. Kimble, Nature {\bf
452}, 67 (2006).

\bibitem{s15}
J. B. Pendry, Phys. Rev. Lett. {\bf 69}, 2772 (1992).

\bibitem{s16}
M. Jachura, M. Karpinski, C. Radzewicz, and K. Banaszek, Opt.
Express {\bf 22}, 8624 (2014).

\bibitem{s17}
D. A. Miller, Nature Photonics {\bf 4}, 3 (2010).

\bibitem{s18}
Kamal, A., Clarke, J. Devoret, M. H, Nature Physics, {\bf 7}, 311
(2011).

\bibitem{s19}
Fan, L., Science, {\bf 335}, 447 (2012).

\bibitem{s20}
Yu Z., Fan S., Nature Photon. {\bf 3}, 91 (2009).

\bibitem{s21}
J. Zi, J. wan and C. Zhang, Appl. Phys. Lett, {\bf 73}, 2084 (1998).

\end{thebibliography}
\end{document}